\documentstyle[12pt]{article}

\begin{document}
\begin{center}
{\Large Self-energy correction to the hyperfine structure splitting of
the $1s$ and $2s$ states in hydrogenlike ions}
\end{center}

\centerline{V.A.Yerokhin$^{1}$, V.M.Shabaev$^{2}$, and A.N.Artemyev$^{2}$}

\begin{center}
$^{1}${\it Institute for High Performance Computing and Data
Bases, Fontanca 118, St.Petersburg 198005, Russia}\\
{\it e-mail: yerokhin@snoopy.phys.spbu.ru}\\

$^{2}${\it Department of Physics, St.Petersburg State University,
 Oulianovskaya 1, Petrodvorets, St.Petersburg 198904, Russia}
\end{center}

\begin{abstract}
The one-loop self-energy correction to the hyperfine structure splitting
of the $1s$ and $2s$ states of hydrogenlike ions is calculated both for the
point and finite nucleus. The results of the calculation are combined
with other corrections to  find the ground state hyperfine splitting
in lithiumlike  $^{209}Bi^{80+}$ and  $^{165}Ho^{64+}$.
\end{abstract}

The recent experimental investigation of  the ground-state
hyperfine splitting of $^{209}Bi^{82+}$ \cite{Klaft} and
$^{165}Ho^{66+}$ \cite{Crespo} shows that the present experimental
accuracy is much higher  than the accuracy of the corresponding
 theoretical values. At present,  measurements
of the ground state hyperfine splitting of lithium-like ions are designed.
In this connection a necessity of an accurate calculation of
the QED corrections to the hyperfine splitting of the $1s$ and $2s$
states of highly charged ions is obvious.

The one-loop self-energy correction to the first-order hyperfine
interaction for the ground state of hydrogenlike ions
in the case of an extended nucleus
was calculated in \cite{Persson} in a wide interval $Z$.
In the case of $Z=83$ and a point nucleus such a calculation
was done in \cite{Shabaev}.
In the present
work we recalculate the self-energy correction for
the $1s$ state and present
results for the $2s$ state.
 The calculation was made using the full-covariant
scheme based on an expansion of the Dirac-Coulomb propagator in terms of
interactions with the external potential \cite{Snyderman,Schneider_PhD}.

The self energy contribution to the hyperfine splitting
is defined by the diagrams shown in Fig.1 where the dotted
line denotes the hyperfine interaction. The formal
expressions for these diagrams can easily be derived
by the two-time Green function method \cite{Shabaev2}.
 The diagrams in Fig.1a are conveniently divided into irreducible
and reducible parts. The reducible part is the part in which the
intermediate-state energy (between the self energy and the hyperfine
interaction line) coincides with the initial-state energy. The irreducible
part is the remaining one. The irreducible part is calculated in the same
way as the first order self-energy contribution. For a point nucleus
the external wave function containing the hyperfine interaction line is
calculated analytically by using the generalized virial relations for the
Dirac-Coulomb problem  \cite{Shabaev3}. For an extended nucleus a
calculation of the external wave function was performed using the reduced
Green function.

The reducible part is grouped with the vertex part presented in Fig.1b.
According to the Ward identity the counterterms for the vertex and the
reducible parts cancel each other and, so, the sum of these terms
 regularized in
the same covariant way is ultraviolet finite. To
cancel the ultraviolet divergences we separate free propagators from
the bound electron lines and calculate them
 in the momentum representation.
The remainder is ultraviolet finite but contains infrared divergences,
which are explicitly separated and cancelled.

The calculations were carried out  for both point and extended nucleus.
In the last case the model of an uniformly charged shell with the radius
$R = \frac{\sqrt{15}}{4} \langle r^2 \rangle^{1/2}$ was used for the nuclear
charge
 distribution.
With high precision, this model is equivalent to the model
of an uniformly charged sphere with the radius
  $R =\sqrt{\frac53} \langle r^2\rangle^{1/2}$
if the first-order hyperfine structure splitting is calculated. Our test
calculation shows a good agreement between these models for the self-energy
correction to the hyperfine splitting too. The Green function
expressed in terms of the Whittaker and Bessel functions \cite{Gyulassy}
was used in the numerical calculation in the case of an extended nucleus.
A part of the vertex term was calculated using the B-spline basis set
method for the Dirac equation \cite{Johnson}.

The results of the calculation for the $1s$ state are listed in the table 1.
The values of the root-mean-square nuclear charge radii given in the
second column of the table are taken
from \cite{Fricke}. The values $X$ listed in the table are defined
by the equation
\begin{eqnarray}
\Delta E_{SE} =\alpha X \Delta E_{nr} \,,
\end{eqnarray}
where $\Delta E_{SE}$ is the self-energy correction to the hyperfine
splitting, $\Delta E_{nr}$ is the non-relativistic value of the hyperfine
splitting (Fermi energy), $\alpha $ is the fine structure constant.
The results of the calculation for a point
nucleus are listed in the third column of the table.
The finite nuclear size contributions and the total self-energy corrections
are given in the forth and fifth  columns, respectively.
 In the last column the values $F$ defined by the equation
\begin{eqnarray}
\Delta E_{SE} = \frac{\alpha}{\pi} F \Delta E_{rel}
\end{eqnarray}
are given.
Here $\Delta E_{rel}$ is the relativistic value of the first-order
 hyperfine
splitting including the finite nuclear size correction. The value
 $F$ is  more stable  than $X$ as respects to a variation of the nuclear
parameters.
In the table 2 the corresponding values for the $2s$ state of hydrogenlike
ions are listed. The relative precision of the results is
estimated to be not worse than $5*10^{-3}$.

As is mentioned above, the self-energy correction was
calculated for the $1s$ state earlier in \cite{Persson, Shabaev}. In
\cite{Shabaev} for $Z=83$ and a point nucleus it was found $X=-3.8$,
while the present calculation gives $X=-3.94$. This discrepancy
results from a noncovariant regularization procedure using
 in \cite{Shabaev},
which, as it turned out, gives a small additional spurious term
\cite{Schneider2}.
 A comparison of the present results for an extended
nucleus with the previous calculation of \cite{Persson} reveals some
 discrepancy too.
So, for $Z=83$ in \cite{Persson} it was obtained  $F=5.098$, while
the present calculation gives $F=5.141$ (a difference due to a discrepancy
between the nuclear parameters is negligible).  A
 detailed comparison of our calculation with one from \cite{Schneider_PhD}
 shows that this discrepancy results from a term in the vertex
 contribution omitted in \cite{Schneider_PhD}.

Taking into account the present results for the self-energy correction and
values of the other corrections (nuclear magnetization distribution,
interelectronic interaction, and vacuum polarization) calculated in
\cite{Shabaeva}, we find that the wavelength of the hyperfine splitting
transition for the $2s$ state in lithiumlike
 $^{209}Bi^{80+}$ and $^{165}Ho^{64+}$ is
 $\lambda=1.548(9)\,\mu$m
($\mu=4.1106(2)\mu_{N}$ \cite{Raghavan})
and $\lambda=4.059(13)\,\mu$m
($\mu=4.132(5)\mu_{N}$ \cite{Nachtsheim},
\cite{Crespo}), respectively.
The uncertainty of these values is mainly given by the nuclear
magnetization distribution correction.

We thank S.M.Schneider for helpful conversations.
The research described in this publication was made
possible in part by Grant No. 95-02-05571a from
the Russian Foundation for Fundamental Investigations.

\newpage
\begin{table}
\caption{Self-energy correction to the hyperfine splitting of the $1s$ state
 in hydrogenlike ions}
\vspace{0.3cm}
\begin{tabular}{|c|c|l|l|l|l|}\hline
$Z$&$\langle r^2\rangle^{1/2}$&$X_{point}$&$\delta X_{fin}$&
$X_{total}$&$ F_{total}$\\ \hline

  49 & 4.598  &    -1.057  &   0.042   &      -1.015   &   -2.629 \\ \hline
  59 & 4.892  &    -1.496  &   0.096   &      -1.400   &   -3.293 \\ \hline
  67 & 5.190  &    -1.995  &   0.192   &      -1.803   &   -3.856 \\ \hline
  75 & 5.351  &    -2.737  &   0.393   &      -2.344   &   -4.470 \\ \hline
  83 & 5.533  &    -3.940  &   0.850   &      -3.090   &   -5.141 \\ \hline
\end{tabular}
\end{table}

\begin{table}
\caption{Self-energy correction to the hyperfine splitting of the $2s$ state
in hydrogenlike ions}
\vspace{0.3cm}
\begin{tabular}{|c|c|l|l|l|l|}\hline
$Z$&$\langle r^2\rangle^{1/2}$&$X_{point}$&$\delta X_{fin}$&$X_{total}$&$
F_{total}$\\ \hline

49 & 4.598 &  -1.131   &      0.048     &      -1.083   &   -2.580 \\ \hline
59 & 4.892 &  -1.690   &      0.109     &      -1.581   &   -3.284 \\ \hline
67 & 5.190 &  -2.381   &      0.238     &      -2.143   &   -3.892 \\ \hline
75 & 5.351 &  -3.475   &      0.520     &      -2.956   &   -4.569 \\ \hline
83 & 5.533 &  -5.372   &      1.208     &      -4.164   &   -5.321 \\ \hline
\end{tabular}
\end{table}
\newpage

\newcommand{\electronline}{\line(0,1){100}} 
\newcommand{\photarctop}{ 
 =E1=A2=A5=E0=E5=E3
  \multiput(0,0)(10,10){3}{\oval(10,10)[tl]}
  \multiput(0,10)(10,10){2}{\oval(10,10)[br]}
  \multiput(40,0)(-10,10){3}{\oval(10,10)[tr]}
  \multiput(40,10)(-10,10){2}{\oval(10,10)[bl]}
}
\newcommand{\pphotarctop}{ 
  \multiput(10,0)(10,10){2}{\oval(10,10)[tl]}
  \multiput(0,0)(10,10){3}{\oval(10,10)[br]}
  \multiput(0,40)(10,-10){3}{\oval(10,10)[tr]}
  \multiput(10,40)(10,-10){2}{\oval(10,10)[bl]}
}
\newcommand{\photonlineleft}{
 \multiput(0,0)(-16,0){3}{\oval(8,8)[b]}
 \multiput(-8,0)(-16,0){2}{\oval(8,8)[t]}}
\newcommand{\photonlineright}{
\multiput(0,0)(16,0){3}{\oval(8,8)[b]}
\multiput(8,0)(16,0){2}{\oval(8,8)[t]}}

\newcommand{\hyperfine}{
\multiput(0,0)(8,0){4}{\line(1,0){2}}
%
\multiput(29,-1)(1,1){3}{.}
\multiput(29,1)(1,-1){3}{.}
%
}

\newcommand{\pphotarcbottom}{ 
  \multiput(0,0)(-10,10){3}{\oval(10,10)[bl]}
  \multiput(-10,0)(-10,10){2}{\oval(10,10)[tr]}
  \multiput(-10,40)(-10,-10){2}{\oval(10,10)[br]}
  \multiput(0,40)(-10,-10){3}{\oval(10,10)[tl]}
}
\newcommand{\ppphotarctop}{ 
   \multiput(0,0)(10,10){6}{\oval(10,10)[tl]}
   \multiput(0,10)(10,10){5}{\oval(10,10)[br]}
   \multiput(100,0)(-10,10){6}{\oval(10,10)[tr]}
   \multiput(100,10)(-10,10){5}{\oval(10,10)[bl]}
}
\newcommand{\loopbottom}{ 
  \put(0,0){\oval(5,5)[bl]}
  \multiput(0,-5)(0,-10){3}{\oval(5,5)[r]}
  \multiput(0,-10)(0,-10){2}{\oval(5,5)[l]}
  \put(0,-30){\oval(5,5)[tl]}
  \put(-30,-30){\electronline}
}
\newcommand{\ppphotwithlooptop}{ 
  \multiput(-1,0)(10,10){4}{\oval(10,10)[tl]}
  \multiput(-1,10)(10,10){4}{\oval(10,10)[br]}
  \multiput(101,0)(-10,10){4}{\oval(10,10)[tr]}
  \multiput(101,10)(-10,10){4}{\oval(10,10)[bl]}
  \put(50,37){\circle{30}}
}

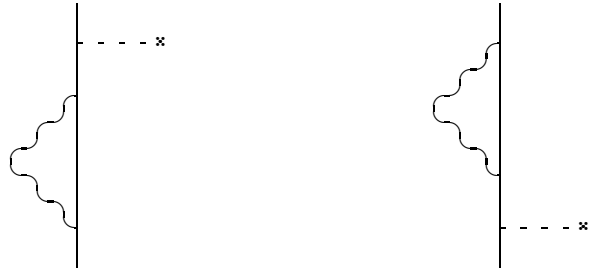
\begin{figure}
\caption{Self energy - hyperfine interaction diagrams.}
  \begin{picture}(480,560)

    \put(140,200){
      \put(40,0){\electronline}
      \put(40,30){\pphotarcbottom}
      \put(40,50){\hyperfine}
}


\put(180,180){b}

    \put(60,410){
      \put(40,0){\electronline}
      \put(40,20){\pphotarcbottom}
     \put(40,85){\hyperfine}
 }

\put(180,390){a}

    \put(220,410){
      \put(40,0){\electronline}
      \put(40,40){\pphotarcbottom}
     \put(40,15){\hyperfine}
}
 \end{picture}
\end{figure}

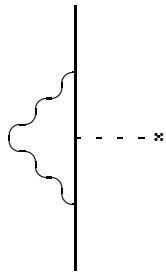
\end{document}